\documentclass[10pt,twocolumn]{article}
\usepackage[letterpaper,left=18mm,right=18mm,top=18mm,bottom=20mm]{geometry}
\usepackage{amsmath,amssymb,graphicx,url}
\usepackage[numbers,sort&compress]{natbib}
\usepackage[font=small,labelfont=bf,labelsep=period]{caption}
\usepackage{microtype}
\newcommand{\authorline}{Yi Yang\\[3pt]
  {\small Institute of Physics, Academia Sinica, Taipei, Taiwan}\\
  {\small Department of Physics, National Cheng Kung University, Tainan, Taiwan}\\
  {\small \texttt{yiyang429@as.edu.tw}}}
\renewcommand{\thetable}{\Roman{table}}

\newcommand{\papertitle}{Positron spectral dips observed near 8 and 20 GV in public AMS-02 data}
\title{\papertitle}
\author{Yi Yang}
\date{}
\begin{document}
\twocolumn[{
\begin{center}
{\Large\bfseries \papertitle\par}\vspace{6pt}
\authorline
\end{center}
\begin{center}\begin{minipage}{.94\textwidth}\small
We project ten years of public AMS-02 daily lepton spectra (2011--2021) onto sky windows using orbital zenith livetime. A dip is a local minimum in the ratio of window-weighted to all-date positron flux as a function of rigidity. Among 1,177 windows centered on Fermi-LAT sources, five show dips at least 3.029 percentage points deep at 7.09--8.48 GV and 34 others at 16.6--22.8 GV. The other four bins contain no dips this deep. For these bins and cut, identified from the data, none of 4,000 within-block date shifts reaches both counts. Allowing any bin pair gives $p=0.0062$ (one-sided $2.50\sigma$) at fixed cut and $p=0.020$ ($2.05\sigma$) after scanning nine cuts. In the upper bin the positron dip exceeds the electron dip at the Galactic center and the Crab by $3.04\pm1.19$ and $3.57\pm1.22$ percentage points, respectively (statistical errors). On a separate sky grid, reselecting dip apertures in every shift and summing the catalog photon fluxes of distinct sources gives an exploratory two-bin tail of $0.0085$ ($2.39\sigma$). With the 58 observed grid apertures fixed, 7 of 360 Galactic-longitude orientations of the catalog yield at least the observed total photon flux when each source is counted once. An isotropic flux with the measured daily spectra reproduces the projections. Testing a celestial origin requires lepton arrival directions.
\end{minipage}\end{center}\vspace{6pt}
}]

The Alpha Magnetic Spectrometer (AMS-02) aboard the International Space Station measures charged cosmic-ray spectra~\cite{amsreview,nuclei}. The positron flux departs from its low-energy power-law trend above $25.2\pm1.8$ GeV, and the positron fraction, $\Phi_{e^+}/(\Phi_{e^+}+\Phi_{e^-})$, rises between 10 and about 250 GeV~\cite{posflux,fraction}. 

The public daily electron and positron fluxes~\cite{electron,positron} and orbital record~\cite{rates} allow us to compare spectral shapes sampled during different viewing periods. Cholis and Hoover searched the AMS positron fraction and lepton fluxes for fluctuations expected from discrete pulsar populations~\cite{cholis}. Milagro found a harder spectrum in one of its TeV cosmic-ray excess regions~\cite{milagro}. Searches based on reconstructed lepton directions constrain anisotropy~\cite{pamela,fermicre,amsang}, including a 95\% credible upper limit of 1.50\% on the positron dipole above 16 GeV~\cite{amsang}.

We weight each direction-integrated AMS spectrum by that day's orbital zenith livetime in windows centered on sources in the Third Catalog of Hard Fermi-LAT Sources (3FHL)~\cite{3fhl}. At a common depth threshold, positron minima recur across several windows in two of the six rigidity bins. We test this recurrence with date shifts and compare the dip locations with catalog photon fluxes on a separate sky grid.

\begin{figure*}[t]
\includegraphics[width=\textwidth]{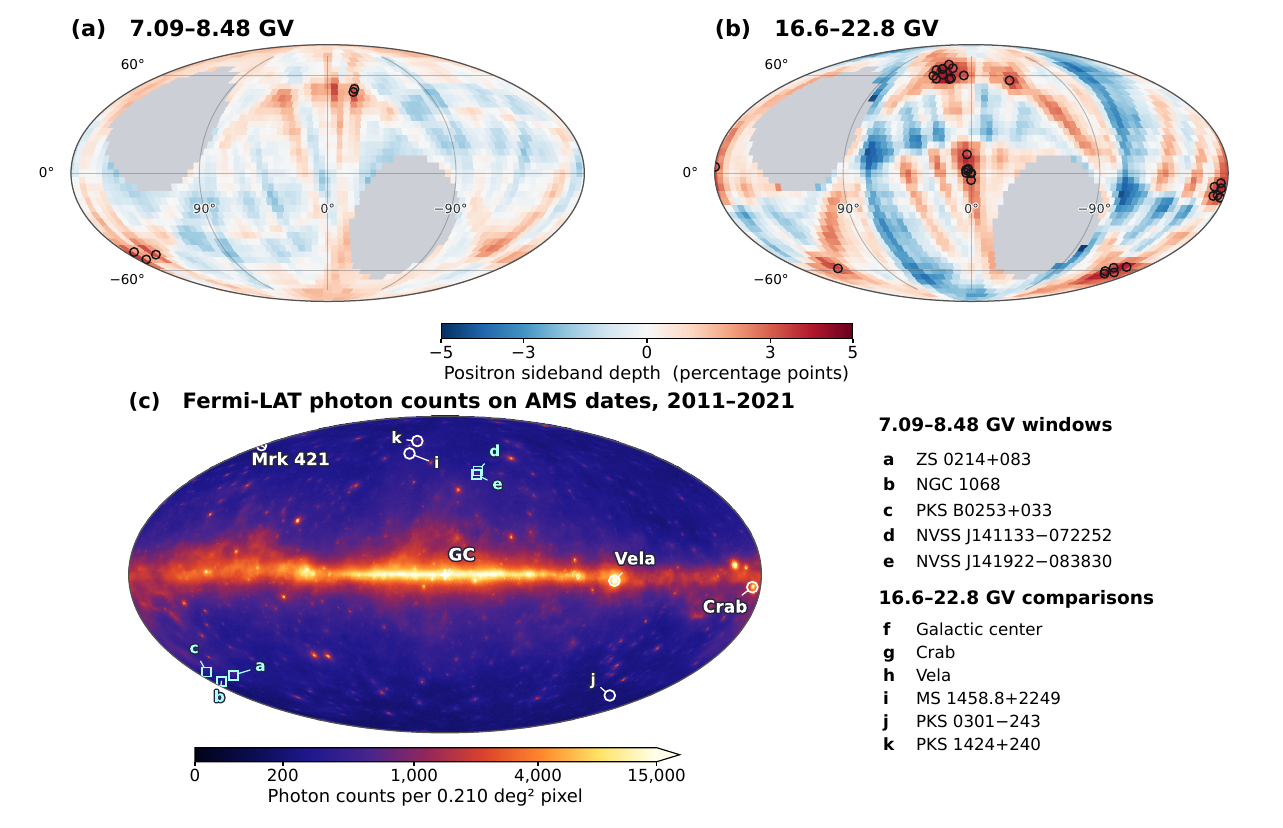}
\caption{(a,b) Positron sideband depth $V_+$ at 7.09--8.48 and 16.6--22.8 GV on a common $\pm5$ percentage-point scale. Circles mark all 5 and 34 catalog centers passing the cut, respectively. Gray marks insufficient AMS exposure. (c) Fermi-LAT photon counts above 100 MeV on AMS observing dates, including diffuse emission. The letters identify the selected source positions. All three maps use Galactic coordinates with longitude increasing leftward. Electron maps and the 3FHL aperture-flux template appear in Figs.~S6 and S7.}
\label{fig:sky}
\end{figure*}

\textit{Orbital projection and dip estimator.} We use eight rigidity bins spanning 2.97-41.9 GV. Let $F_{skd}$ denote the published flux of species $s$ (positron or electron) in rigidity bin $k$ on UTC date $d$. The two series have 3268 common dates from May 20, 2011, to November 2, 2021. For each date, AMS02RATES supplies spacecraft position and livetime. We project the geocentric radial direction onto the sky. The resulting orbital weights are common to both charges and all rigidities.

We label a window by $a$ and its center by Galactic longitude $l$ and latitude $b$. Each window spans $8^\circ\times16.36^\circ$ in local longitude and latitude, with its long axis pointing toward Galactic north. One position per populated minute represents its accepted livetime. For livetime $L_{dah}$ in the northern or southern half $h$ on date $d$, we average the normalized daily weights, $w_{da}=\tfrac12\sum_h L_{dah}/\sum_{d'}L_{d'ah}$, so that unequal exposure on the two sides does not favor one half. For the reference, $g_d$ is the day's share of accepted livetime over the entire orbit. Both use the full date sample~\cite{sm}.

The percent deviation of the window-weighted flux from the all-date reference for the same species and rigidity is
\begin{equation}
 M_{sk}(a)=100\left[\frac{\sum_d w_{da}F_{skd}}{\sum_d g_d F_{skd}}-1\right].
 \label{eq:m}
\end{equation}
We interpolate the two sideband deviations in $\log R$ and subtract the deviation in the central bin,
\begin{align}
 V_{sk}(a)&=(1-u_k)M_{s,k-1}(a)+u_kM_{s,k+1}(a)-M_{sk}(a),\notag\\
 D_k(a)&=V_{+k}(a)-V_{-k}(a).
 \label{eq:depth}
\end{align}
Here $u_k=\ln(R_k/R_{k-1})/\ln(R_{k+1}/R_{k-1})$ and $R_k$ is the geometric bin center. The 7.09--8.48 GV bin uses the 5.90--7.09 and 8.48--11.0 GV sidebands. The 16.6--22.8 GV bin uses 11.0--16.6 and 22.8--41.9 GV. A positive $D_k$ means a deeper relative positron valley, in percentage points (pp).

For fixed $k$, normalize each daily flux to its all-date mean, $y_{sjd}=F_{sjd}/\sum_{d'}g_{d'}F_{sjd'}$. At $R_k$ the line through the two sidebands in $\ln R$ predicts $\alpha_{sd}=(1-u_k)y_{s,k-1,d}+u_k y_{s,k+1,d}$. Subtracting the central value cancels a common offset and a linear tilt. The daily charge difference in dip depth is
\begin{equation}
 q_{kd}=100\big[(\alpha_{+d}-y_{+kd})-(\alpha_{-d}-y_{-kd})\big].
 \label{eq:daily}
\end{equation}
A dip in the normalized daily positron flux accompanied by a flat normalized electron spectrum gives $q_{kd}>0$. A fixed difference in the mean spectral indices cancels in the bin normalization. A varying index can contribute through curvature of the normalized flux in $\ln R$.

\begin{figure*}[t]
\centering
\includegraphics[width=.99\textwidth]{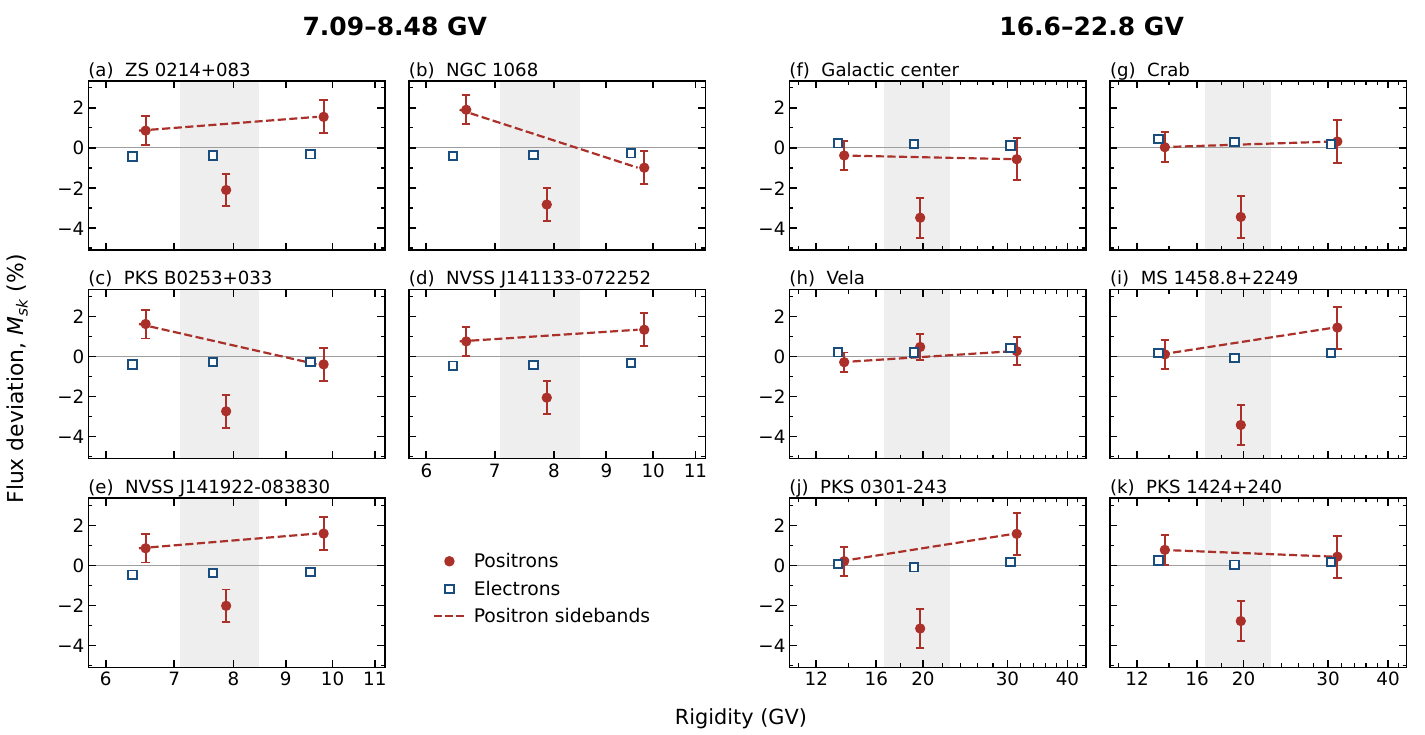}
\caption{Projected flux deviations for (a--e) all five catalog windows passing the 7.09--8.48 GV cut and (f--k) six comparison windows at 16.6--22.8 GV, including Vela. Red circles (positrons) and open blue squares (electrons) show statistical errors. Points are displaced horizontally for visibility. The dashed line interpolates the positron sidebands in $\ln R$. Its value minus the central-bin measurement (shaded bin) gives $V_+$.}
\label{fig:shape}
\end{figure*}

\textit{Six-bin rigidity scan.} At all 1177 catalog centers we require a central positron offset below both neighbors and a depth $V_{+k}$ at least as large as the 16.6--22.8 GV Galactic-center (GC) depth (3.029 pp, rounded). In ascending rigidity the six counts are 0, 0, 5, 0, 0 and 34. Figure~\ref{fig:sky}(a,b) marks all passing centers on the two positron depth maps. Figure~\ref{fig:sky}(c) locates selected sources on the full Fermi-LAT photon sky above 100 MeV on AMS observing dates~\cite{latdata}. The five windows at 7.09--8.48 GV have depths of 3.08--3.45 pp [Fig.~\ref{fig:shape}(a--e)]. The median depths in the two bands are 3.28 and 3.46 pp. No center passes in both bands, although neighboring apertures can overlap. To calibrate these counts, we cyclically shift the spectra by a random number of observed dates within each 27-day Bartels block. All charges, rigidities and windows share the same offset within a block. Each block is shifted independently, with orbital weights fixed. At this cut, 104 of 4,000 date shifts yield at least five passing windows in the lower bin, and 83 yield at least 34 in the upper bin. Including the observation in the empirical ranks gives fixed-bin tails of 0.0262 and 0.0210. Allowing any of the six bins to yield a count with an equally small binwise tail gives 0.119 and 0.091, respectively~\cite{sm}. Over the eligible grid, electron depths range from $-0.41$ to $+0.36$ pp in the lower band and $-0.44$ to $+0.48$ pp in the upper band (Fig.~S6).

With the observed bins fixed retrospectively, none of the 4,000 shifts reaches both counts. We rank each count against shifts in its own bin and take the second-smallest tail across bins, requiring at least two windows per contributing bin. At the GC-derived cut, 24 of 4,000 shifts match or exceed the observed recurrence in any two bins ($p=25/4001=0.0062$, one-sided $2.50\sigma$). Scanning nine cuts from 2.00 to 4.00 pp gives the strongest observed recurrence at 3.25 pp (three lower-bin and 23 upper-bin windows). With the same search applied to every shift, 80 shifts match or exceed the observed recurrence statistic ($p=81/4001=0.020$, one-sided $2.05\sigma$)~\cite{sm}. The recurrence statistic was chosen after inspecting the spectra. This calibration holds the catalog, apertures and dip definition fixed.

\textit{Galactic-center and Crab projections.} The separate 16.6--22.8 GV grid search for a pair of regions required a negative central positron offset below both sidebands and electron curvature consistent with zero. Starting at the GC, we chose a partner patch by overlap with the antipodes of the GC patch and took its deepest positron valley~\cite{sm}. Its grid center is $(l,b)=(-176^\circ,-8^\circ)$ near the Galactic anticenter. This is $2.29^\circ$ from the Crab. In the timing test described below, 1261 maps met the observed GC shape and minimum positron depth after the dates of the spectra were shifted. Of these, 74 (5.9\%) gave a partner at least this close to the Crab under the same rule. A closer catalog object, TXS~0518+211, is $0.81^\circ$ away, but the Crab accounts for 83.5\% of the modeled point-source photon flux in this window. We evaluate the lepton estimator at the mathematical GC, $(0^\circ,0^\circ)$, and at the 3FHL Crab position, $(184.55^\circ,-5.78^\circ)$, as shown in Fig.~\ref{fig:shape}(f,g). The central positron deviations are $(-3.49\pm1.00)\%$ for the GC and $(-3.45\pm1.03)\%$ for the Crab. At both positions the positron sidebands are within $0.6\%$ of their references. Table~\ref{tab:measure} lists $V_+$, $V_-$ and their difference for 17 reference, grid and catalog positions at this rigidity. This is the highest bin with an upper sideband, so the high-rigidity width is unresolved.

\begin{table*}[t]
\caption{Projected depths at 16.6--22.8 GV in percentage points, with statistical errors. $D=V_+-V_-$. The left block gives reference and catalog positions. The right block gives the selected partner grid center $(184^\circ,-8^\circ)$, its nearest catalog source, and the three photon-bright regions with their principal sources (Fig.~S7). The numerical tables cover all 1177 eligible catalog centers.}
\label{tab:measure}
\centering\small\setlength{\tabcolsep}{2.6pt}
\begin{tabular}{lrrr@{\hspace{2em}}lrrr}
Position & $V_+$ & $V_-$ & $D$ & Position & $V_+$ & $V_-$ & $D$\\\hline
Galactic center & $3.03\pm1.18$ & $-0.01\pm0.16$ & $3.04\pm1.19$ & Partner grid center & $3.86\pm1.22$ & $0.08\pm0.16$ & $3.78\pm1.23$\\
Crab & $3.61\pm1.21$ & $0.04\pm0.16$ & $3.57\pm1.22$ & TXS 0518+211 & $3.73\pm1.22$ & $0.08\pm0.16$ & $3.65\pm1.23$\\
Vela & $-0.52\pm0.78$ & $0.10\pm0.10$ & $-0.62\pm0.78$ & Region 1 & $1.49\pm1.28$ & $0.00\pm0.17$ & $1.49\pm1.30$\\
Mrk 421 & $-0.06\pm1.01$ & $-0.10\pm0.13$ & $0.04\pm1.02$ & Region 2 & $-1.08\pm0.94$ & $-0.13\pm0.12$ & $-0.95\pm0.95$\\
MS 1458.8+2249 & $4.12\pm1.17$ & $0.26\pm0.15$ & $3.86\pm1.18$ & Region 3 & $-2.18\pm1.10$ & $-0.06\pm0.14$ & $-2.12\pm1.11$\\
PKS 0250$-$225 & $4.11\pm1.18$ & $0.23\pm0.16$ & $3.88\pm1.19$ & PG 1553+113 & $0.90\pm1.26$ & $-0.02\pm0.16$ & $0.92\pm1.27$\\
PKS 0301$-$243 & $3.97\pm1.15$ & $0.22\pm0.15$ & $3.75\pm1.16$ & 3C 66A & $-0.80\pm0.74$ & $-0.19\pm0.10$ & $-0.61\pm0.75$\\
PKS 1424+240 & $3.41\pm1.18$ & $0.17\pm0.15$ & $3.24\pm1.19$ & B3 0133+388 & $-0.83\pm0.88$ & $-0.20\pm0.11$ & $-0.63\pm0.89$\\
 & & &  & PKS 2155$-$304 & $-0.06\pm1.09$ & $-0.06\pm0.14$ & $0.00\pm1.10$\\
\end{tabular}

\end{table*}

\textit{Source positions and photon brightness.} On the 2,094 eligible AMS grid centers, the 3.029 pp cut selects 8 lower-bin and 50 upper-bin dip apertures. To compare their locations with the photon sky, we hold all 58 apertures fixed and rotate the entire 3FHL catalog in Galactic longitude in $1^\circ$ steps, preserving source latitudes and separations. Every aperture contains a source in 216 of 360 orientations, including the observed one. Summing the 16.6--22.8 GeV catalog photon flux of each distinct source once across the 58 apertures (Fig.~S7), only 7 of 360 orientations reach or exceed the observed flux~\cite{3fhl,sm}.

For the timing test we keep the catalog fixed and reselect the dip apertures on every date-shifted map. We use the same photon band at every rigidity and count each source once per rigidity bin. At the 3.029 pp cut the two observed groups cover 31 and 102 distinct sources, respectively, with no shared entry. The timing statistic depends on both the sky area selected in each bin and its photon brightness. With any bin pair and nine depth cuts, 33 of 4,000 shifts reach the observed weighted recurrence ($p=34/4001=0.0085$, one-sided $2.39\sigma$). The unweighted count on this same grid gives $p=0.045$. Pooling distinct sources over all six bins gives $p=35/4001=0.0087$ (one-sided $2.38\sigma$) for the total flux~\cite{sm}.

We compare charge-dependent dip depths at fixed Galactic latitude. For aperture flux $G(a)$, latitude-ring mean $\bar G_{b(a)}$ and grid-cell solid angle $\Omega_a$, define $h_a=\Omega_a[G(a)-\bar G_{b(a)}]$. Positive and negative weights balance on each ring, so $T_k=\sum_a h_a D_k(a)/\sum_{h_a>0}h_a$ measures the weighted difference. At 16.6--22.8 GV, $T_k=0.573\pm0.228$ pp, with separate positron and electron contrasts of $0.560\pm0.226$ and $-0.012\pm0.029$ pp, respectively. The other five interior bins give $-0.062$ to $+0.052$ pp. After excluding windows that could overlap the GC, Crab or their antipodes and recalculating each ring mean, the remaining 1932 centers give $0.151\pm0.197$ pp. Evidence for a brightness trend away from these regions is weak.

Vela [Fig.~\ref{fig:shape}(h)] shows little curvature. Vela and the Crab have fitted fluxes of 1.539 and 1.503 in units of $10^{-9}\,{\rm cm}^{-2}\,{\rm s}^{-1}$, yet $D_{\rm Crab}-D_{\rm Vela}=4.19\pm1.52$ pp with shared covariance. This contrast is conditional on the selected Crab position and rigidity. Third-ranked Mrk~421 gives $D=0.04\pm1.02$ pp. The GC flux sums 14 point sources and excludes diffuse emission.

Centering regions 1--3 in Fig.~S7 on their dominant photon sources, PG~1553+113, 3C~66A and PKS~2155$-$304, gives $V_+=+0.90\pm1.26$, $-0.80\pm0.74$ and $-0.06\pm1.09$ pp. A scan centered on all 3FHL point sources gives 1177 eligible positions. The two deepest apertures are nearly antipodal. The largest positron depth is $4.12\pm1.17$ pp at MS~1458.8+2249 [Fig.~\ref{fig:shape}(i)]. PKS~1424+240 dominates its photon flux, and PKS~0301$-$243 dominates the second aperture. Centered on PKS~0301$-$243 and PKS~1424+240, the depths are $3.97\pm1.15$ and $3.41\pm1.18$ pp, see Fig.~\ref{fig:shape}(j,k). Their statistical correlation is $0.486$. These follow-ups use the same map that selected the targets. At fixed rigidity, 13.8\% of trials with shifted dates give a source-centered maximum at least as deep as observed~\cite{sm}.

\textit{Cross-checks.} Across eligible grid centers, the unweighted Pearson correlation of $V_+$ with $D$ at 16.6--22.8 GV is 0.9955 (difference map in the Supplemental Material~\cite{sm}). Pooling the two halves in proportion to their livetime gives $D=3.04\pm1.19$ pp (GC) and $3.54\pm1.22$ pp (Crab). Quoted uncertainties propagate the published statistical errors through numerator and denominator, treating dates, bins and species as independent inputs and retaining their shared contributions across windows. At both anchors, the positron terms account for 98.3\% of the propagated statistical variance of $D$. A full systematic covariance is unavailable.

The GC lies $7.35^\circ$ from the Crab antipode. Of the 501 GC and 467 Crab dates, 350 are shared and carry 80.8\% and 84.0\% of the respective weights. The correlation between their charge differences $D$ is 0.680. Including this covariance gives $D_{\rm Crab}-D_{\rm GC}=0.53\pm0.96$ pp. Omitting the shared dates from both targets and the reference gives $3.41\pm2.67$ and $5.83\pm2.81$ pp. Recomputing the two exposures from one-second orbital positions gives $3.05\pm1.19$ and $3.58\pm1.22$ pp, respectively, on the same full sample. 

Splitting the data at January 1, 2016 gives GC and Crab charge depths of $(5.38,5.17)$ pp in the first period and $(0.76,2.08)$ pp in the second. Omitting any ten consecutive calendar days leaves at least 2.67 and 3.23 pp. Omitting the 20 dates with the largest absolute contributions in each projection leaves 1.29 and 1.93 pp. This data-selected deletion measures the influence of individual dates. Its conditional calibration and the period uncertainties are given in the Supplemental Material~\cite{sm}.

For the GC and Crab, we use the same block-shift procedure. To require a departure in both windows in the same bin, we take the smaller of their depths after standardizing each by its null mean and dispersion~\cite{sm}. With both positions fixed, the pair timing tail is 0.00435 at 16.6--22.8 GV and 0.02385 when any of six bins can supply the pair. These probabilities quantify timing conditional on the selected positions. For the photon-weighted contrast $T_k$, the six-bin tail is 0.0166 with the nominal block origin and 0.0395 or 0.0685 for 9- or 18-day block phases. Rotating the catalog in Galactic longitude instead gives a contrast at least as large in 27 of 360 orientations. 

We also rotated the two apertures rigidly together around either the Galactic or equatorial pole. At a common right-ascension shift of $146^\circ$, the centers move to Galactic latitudes $-55.2^\circ$ and $+56.3^\circ$. Their depths are $3.08$ and $3.23$ pp with only $30\%$ and $2.3\%$ of the respective original photon fluxes. The rotated windows contain PKS~0301$-$243 and MS~1458.8+2249, respectively. The two scans share sky structure. Scanning both paths at the target rigidity gives a pair score at least as large as the observed score in $37.6\%$ of the maps reconstructed from the shifted daily spectra. Allowing any of the six rigidities along either path raises this fraction to $92.0\%$. Requiring both windows to be at least as photon-bright as the original pair gives $0.85\%$ at fixed rigidity and $4.62\%$ across six bins. Only four distinct orientations pass these brightness thresholds, all overlapping the original pair. This conditional comparison cannot calibrate the initial choice of pair, geometry, photon thresholds or observable. No discovery-level probability is assigned.

\textit{Physical interpretation.} Since $\sum_d g_d y_{sjd}=1$ for every bin and charge, the all-date mean of $q_{kd}$ vanishes [Eq.~(\ref{eq:daily})]. Thus $D_k(a)=\sum_d(w_{da}-g_d)q_{kd}$ measures how the orbital weights align with changes in daily spectral shape. Varying exposure to a steady directional dip could produce such changes. An isotropic sky with the measured spectrum on each date also reproduces every projection exactly. AMS reports charge-dependent time variations~\cite{positron}. The daily tables alone cannot separate temporal and celestial contributions.

Can nonnegative local intensities produce projected depths this large? A zenith pencil beam retains 2.41\% (GC) or 2.88\% (Crab) of a small deficit confined to its own aperture. Cross-response between the apertures also matters. With both responses and the reference normalization included, a 75\% central-bin positron deficit in each aperture gives depths of 3.25 and 3.46 pp~\cite{sm}. The daily projection strongly dilutes a local signal under this angular response.

Could in-flight annihilation produce the observed rigidity dependence? For a 20 GV positron on an electron at rest the two-photon cross section is $6.55\times10^{-29}\,\mathrm{cm^2}$~\cite{annih}. A column giving 75\% attenuation at 20 GV removes 86\% and 61\% at the lower and upper sideband centers. The projected positron depths are only 0.040 and 0.042 pp. In the same two-window pencil-beam model the depth stays below 0.10 pp at either position for \emph{every} common column~\cite{sm}. The smooth rigidity dependence of direct annihilation strongly suppresses the sideband contrast.

\textit{Conclusions.} In the six-bin scan, the selected positron minima recur at 7.09--8.48 GV in five catalog-centered windows and at 16.6--22.8 GV in 34 others. At the observed bins and cut, none of 4,000 date shifts reaches both window counts. Reselecting the bin pair and cut in each date shift gives exploratory two-bin tails of $0.020$ for catalog-centered window counts and $0.0085$ for the distinct-source photon flux on the AMS grid. With the observed grid apertures fixed, only 7 of 360 catalog orientations in Galactic longitude give as much total distinct-source flux. These timing and angular comparisons use the same observed dip map. In the upper band, the GC and Crab projections retain a dip under alternative livetime weights, while Vela, with a fitted photon flux comparable to the Crab's, shows little curvature. Direct annihilation on a common cold-electron column gives too little curvature in the two-window model. An isotropic flux with the measured time dependence reproduces both bands in orbital projection. With reconstructed arrival directions, a comparison of target and control windows on the same dates could test whether the dips recur at these rigidities on the sky~\cite{sm}.

\noindent\textit{Acknowledgments.}
I thank the AMS Collaboration for making the daily electron and positron flux tables publicly available, NASA's High Energy Astrophysics Science Archive Research Center for maintaining the AMS02RATES orbital and livetime records, and the Fermi-LAT Collaboration for releasing photon and spacecraft data and the 3FHL catalog.
I gratefully acknowledge support from Academia Sinica, National Cheng Kung University, and the National Science and Technology Council (NSTC) of Taiwan.

\textit{Data availability.} The flux tables, orbital records, LAT event files and 3FHL catalog are public~\cite{electron,positron,rates,3fhl,latdata}. The source code, reduced inputs and numerical results are available from the author upon reasonable request. 
\textit{AI use disclosure.} I used OpenAI ChatGPT and Codex to assist with the analysis code, numerical checks, figures, literature review, and drafting. I take responsibility for the calculations, their interpretation, and the final text.

\onecolumn
\setcounter{figure}{0}
\setcounter{table}{0}
\setcounter{section}{0}
\setcounter{equation}{0}
\begin{center}
{\large Supplemental Material for\par}\vspace{3pt}
{\Large\bfseries \papertitle\par}\vspace{6pt}
\authorline
\end{center}

\renewcommand{\thetable}{S\arabic{table}}
\renewcommand{\thefigure}{S\arabic{figure}}
\section{Public data and geometric sampling}
We use the AMS electron tables S1-S3300 and positron tables S1-S3268, released with Phys. Rev. Lett. \textbf{130}, 161001 (2023) and \textbf{131}, 151002 (2023), the NASA HEASARC AMS02RATES daily all-particle FITS files, and version 13 of the Third Catalog of Hard Fermi-LAT Sources (3FHL), distributed by the Fermi Science Support Center. Re-downloaded AMS CSV files and the LAT FITS file were byte-identical to the inputs used for the initial scan. Independent parsing gave exactly the same flux and statistical error in all $3268\times8\times2=52\,288$ matched entries. Rigidity bin edges are 2.97, 4.02, 5.90, 7.09, 8.48, 11.0, 16.6, 22.8, and 41.9 GV. Dates are indexed in UTC.

For the orbital geometry, the minute reduction requires latitude in $[-90^\circ,90^\circ]$, $|\mathrm{longitude}|\leq360^\circ$, $0<L\leq1.00001$ s of livetime, $6500<r<7100$ km, and a tabulated zenith angle in $[0^\circ,40^\circ]$. These are analysis cuts on the all-particle records. Neither that angle nor the event rate gives a reconstructed lepton direction. The exact-second target check also requires nonnegative longitude. In each minute with accepted livetime we use the valid record nearest second 30, choosing the earlier record in a tie. We evaluate its direction at the center of that one-second interval and assign it the sum of the accepted livetimes in the minute. We rotate the geocentric radial vector using Greenwich mean sidereal time (UTC as an approximation to UT1), precess to J2000, and transform to Galactic coordinates.

For unit direction $\boldsymbol v$ and the local center, east and north unit vectors $(\boldsymbol c,\boldsymbol e,\boldsymbol n)$, the aperture requires
\begin{equation}
 |\mathrm{atan2}(\boldsymbol v\!\cdot\!\boldsymbol e,\boldsymbol v\!\cdot\!\boldsymbol c)|<4^\circ,\qquad
 |\arcsin(\boldsymbol v\!\cdot\!\boldsymbol n)|<90^\circ/11.
\end{equation}
The sign of $\boldsymbol v\!\cdot\!\boldsymbol n$ selects the northern or southern half of the aperture in the local frame. The center coordinates $(l,b)$ are Galactic longitude and latitude. Each half needs $(\sum_dL_d)^2/\sum_d L_d^2\geq100$ effective dates for the original scan. The 2584 centers use $4^\circ$ latitude spacing between $-88^\circ$ and $+88^\circ$ and two polar centers. On each nonpolar ring the number of uniformly spaced longitudes is $\max[2,2\,\mathrm{round}(45\cos b)]$. Cell areas are the spherical ring areas between adjacent latitude midpoints divided by the number of centers on the ring.

As an independent check of the minute approximation, we also evaluate the GC and Crab apertures from one-second RATES positions. A daily sky histogram made from individual seconds in $2^\circ\times2^\circ$ equatorial pixels screens for dates that could intersect either footprint. The screen dilates the aperture corner radius by $1.5^\circ$, exceeding the maximum distance of a point in a pixel from its center. The screen serves only to identify zero-exposure dates. For candidate dates, cuts and aperture boundaries are applied directly to the individual seconds.

All 864 candidate daily FITS files passed their archived SHA-256 checksums, and their accepted row counts and total livetimes matched the daily manifest. The other 2404 common dates had no accepted second in a candidate pixel. The direct calculation gives GC and Crab charge depths of $3.05\pm1.19$ and $3.58\pm1.22$ pp. Nonzero exposure occurs on 525 and 487 dates, including 370 shared dates. When all 370 shared days of the one-second reconstruction are removed and all weights renormalized, the depths become $3.23\pm2.69$ and $6.01\pm3.22$ pp. An independently implemented test with 150\,000 synchronous shifts gives one-sided fixed-bin and six-bin probabilities of 0.0044 and 0.0244 with the one-second weights, compared with 0.0046 and 0.0252 with the minute weights and the same random offsets. These comparisons test time sampling and reuse the same AMS fluxes. They provide no independent event sample. The catalog and rotation scans retain the common minute prescription across the sky. An unscreened replay from freshly downloaded FITS files on all 3268 dates verified each file digest and examined 272\,962\,846 one-second records. All 4\,601\,535 populated minutes reproduced the archived positions, sample offsets and livetimes. Both reconstructed target exposures agreed in every half-window on every date to within $2\times10^{-8}$ s.

\section{Curvature and shared statistical errors}
For charge species $s$, rigidity bin $k$, and UTC date $d$, let $F_{skd}$ denote the published daily flux. With the window date weights $w_{da}$ and all-orbit date weights $g_d$ of the main text, let $B_{sk}=\sum_dg_dF_{skd}$, $A_{ask}=\sum_dw_{da}F_{skd}$, and $H=(1-u,-1,u)$ in ascending rigidity order. The derivative of a bin deviation with respect to its daily flux is
\begin{equation}
 J_{adsk}=\frac{100}{B_{sk}}\left(w_{da}-g_d\frac{A_{ask}}{B_{sk}}\right).
\end{equation}
For fixed $k$, set $y_{sjd}=F_{sjd}/B_{sj}$. Choosing the central rigidity as pivot, $R_*=R_k$, the sidebands determine
\[
 \beta_{sd}=\frac{y_{s,k+1,d}-y_{s,k-1,d}}{\ln(R_{k+1}/R_{k-1})},\qquad
 \alpha_{sd}=y_{s,k-1,d}+\beta_{sd}\ln(R_k/R_{k-1}).
\]
Both coefficients refer to this triplet. Taking the sideband prediction minus the measured central value, first for positrons and then subtracting the same quantity for electrons, gives Eq.~(3) of the main text. Equivalently, $q_{kd}=100\sum_{j=k-1}^{k+1} H_j(y_{+jd}-y_{-jd})$, with $(H_{k-1},H_k,H_{k+1})=(1-u_k,-1,u_k)$. Since each $y_{sjd}$ has unit $g$-weighted mean, $\sum_d g_dq_{kd}=0$ and $D_k(a)=\sum_d(w_{da}-g_d)q_{kd}$. The statistical covariance of $D_k(a)$ and $D_k(b)$ is $\sum_{dsj}H_j^2J_{adsj}J_{bdsj}\sigma_{dsj}^2$, summed over the three bins and two species, under the independent-input statistical approximation. The charge sign appears twice in each same-species covariance term. Keeping only one species gives its two-window covariance. For a difference of two window depths the variance is $C_{aa}+C_{bb}-2C_{ab}$. For the on-sky map each full-aperture estimate uses both halves. Their equal weighting compensates unequal integrated livetime on either side of the center. It does not split the flux sample into two independent data sets. A separate implementation used direct ratios of sums and checked their analytical gradients by two-sided finite variations of the original fluxes. For the minute exposures, deleting the 350 shared dates renormalizes both half-window weights and the reference on the same 2918 dates. For that diagnostic only, effective dates per half fall to $(88,30)$ for the GC and $(69,23)$ for the Crab, below the full-scan cut.

The AMS tables also give per-bin time-dependent systematic errors and total systematic errors. As a sensitivity calculation, treating the published time-dependent components as fully diagonal would add $0.085$ and $0.088$ pp in quadrature to the GC and Crab charge errors, respectively. Those numbers quantify one correlation assumption. The released marginal errors do not determine the covariance between days, charges and rigidities. As a geometric control, using $w_{da}=\sum_hL_{dah}/\sum_{d'h}L_{d'ah}$ yields $D=(3.04,3.54,-0.63,0.09)$ pp for the GC, Crab, Vela and Mrk~421 and a Crab-minus-Vela contrast $4.17\pm1.52$ pp. This combines the full aperture in proportion to livetime. A fixed multiplicative scale for any one charge and bin cancels exactly in its normalized deviation. Time-varying errors need not cancel.

\paragraph{Finite angular deficits.} Assume a zenith pencil beam and uniform steady baseline. Model a fractional positron deficit $x_j\in[0,1]$ in the central rigidity bin only, uniform in each of the two disjoint sky apertures. Keep electrons and the two sidebands fixed. Let $f_{dj}=\sum_hL_{djh}/L_{d,\mathrm{total}}$, $\bar f_j=\sum_dg_df_{dj}$ and $A_{aj}=\sum_d(w_{da}-g_d)f_{dj}$. The daily central flux is proportional to $1-\sum_jf_{dj}x_j$. Normalization to the all-date mean gives the exact depth in this model,
\[
 V_{+}(a)=100\,\frac{(A x)_a}{1-\bar f\cdot x}\quad\mathrm{pp},\qquad
 A=\begin{pmatrix}0.02414&0.01895\\0.01705&0.02880\end{pmatrix}.
\]
Rows and columns follow GC, Crab order, and $\bar f=(0.00307,0.00339)$. The two observed central-bin deviations alone correspond to $x=(0.94,0.64)$, whereas matching the two observed sideband-referenced depths gives $x=(0.50,0.95)$. These are separate inversions of noisy summaries. A common illustrative $x=(0.75,0.75)$ gives 3.25 and 3.46 pp. Removing the 350 shared dates and recomputing $A$ yields rows $(0.01589,-0.00060)$ and $(-0.00067,0.01573)$. The same common deficit then predicts 1.15 and 1.13 pp. For the two calendar periods in the main text this example gives (3.23, 3.43) and (3.26, 3.48) pp, respectively. The measured positron-only depths after deletion are $3.66\pm2.65$ and $5.99\pm2.79$ pp. For a separate check of the single-aperture limit, take the Crab completely dark in the target bin and give the GC its baseline intensity. The predicted positron depths are then $(1.90,2.89)$ pp. Their joint squared residual from the measured $V_+$ values is 0.92 using the propagated two-window statistical covariance. Exact reproduction of both central values and statistical consistency are distinct requirements. The fitted amplitudes and the boundary example illustrate geometric dilution. An angular measurement requires the experimental lepton response.
\section{Fermi template and rigidity controls}
The 3FHL source spectra are seven-year catalog fits, covering August 2008 to August 2015. The source rankings and brightness tests use these time-averaged catalog fluxes. The event map in Fig.~\ref{fig:sky}(c) of the main text uses the AMS observing dates as described below. The catalog contains 1508 entries with empty extended-source names. We integrate the best-fit power law or log parabola in photon energy, using $dN/dE\propto (E/E_0)^{-\Gamma-\beta\ln(E/E_0)}$ for a log parabola and $\beta=0$ for a power law. We multiply the integral flux in the catalog by the ratio of the target-band and normalization-band model integrals, normalizing to the catalog integral \texttt{Flux} on 10-1000 GeV. Integration over the target and normalization bands was independently repeated by adaptive quadrature. The two brightest photon entries differ in nominal fitted 16.6-22.8 GeV flux by 2.4\%. Errors in the fitted photon spectra do not enter the quoted AMS statistical errors.

The GC and Crab windows contain 14 and 5 photon point sources, with total band fluxes $5.005\times10^{-10}$ and $1.799\times10^{-9}\,\mathrm{cm}^{-2}\,\mathrm{s}^{-1}$, respectively. The GC window is centered on $(l,b)=(0,0)$. The scan retained central positron deviations below zero and both sidebands, electron curvatures within 1.96 statistical errors of zero, and positron depths at least as large as the GC depth in that map. For this exploratory selection only, the original electron cut fixed each window width across trials and used $J\simeq100(w-g)/B$ in its statistical error. Restoring $A/B$ in $J$ changes the width in $V_-$ by less than $4.5\times10^{-6}$ pp in every eligible cell and leaves every observed electron pass/fail decision unchanged. The quoted measurement errors use the full derivative. The patch algorithm connected retained grid centers within $6.1^\circ$. The partner patch had the most centers antipodal to members of the GC patch, excluding the GC patch itself. The deepest positron valley broke ties and represented each patch. Photon positions entered only afterward. In 100\,000 shifted skies, 1261 met the GC shape criteria with GC depth at least as large as observed. Of 1014 skies with a partner in this subset, 74 matched the Crab within $2.29^\circ$ (7.3\%). Counting a missing partner as no match gives 74/1261 (5.9\%). The full-template statistic uses area weights and subtracts the mean across eligible windows at the same latitude. Direct ratios of flux sums and an independent catalog-membership calculation reproduce all six contrasts.

\begin{figure}[h]
\includegraphics[width=.50\textwidth]{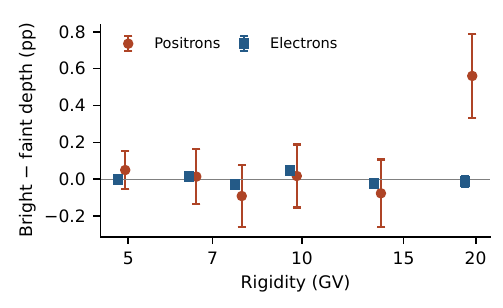}
\caption{Signed contrast for each charge between photon-brighter and photon-fainter windows within the same latitude rings. The six points per charge use the six interior rigidity bins. Errors are statistical.}
\end{figure}

To excise the GC and Crab regions, we discard every candidate window whose center is within twice the aperture corner radius of the GC, the Crab, or either antipode. This $18.20^\circ$ cut guarantees disjoint footprints. We recompute the latitude-ring means and normalize the positive and negative photon weights anew. The surviving $1932$ windows give $T=0.151\pm0.197$ pp, with a six-bin timing tail of $0.443$. Retained and excluded windows still use the same daily fluxes. Removing only the Crab catalog entry gives $T=0.212$ pp, with six-bin timing tail 0.240.

\section{Conditional probability calculations}
Within each 27-day Bartels block we cyclically shift the sequence of \emph{observed} dates by a random offset. Missing days have no entries. All charges, bins and windows in the block share the same offset, and different blocks have independent offsets. BR 1 begins on February 8, 1832. With $n$ the number of calendar days since that date, the block index for phase $m$ is $\lfloor(n+m)/27\rfloor$. We use $m=0,9,18$ days. The grid-based pair, patch and photon-contrast scans keep the all-date bin means fixed and permute the normalized daily fields. Each resulting map is one realization of a null hypothesis in which the phase of the spectra within each block is random relative to the orbit. Time variability, changing measurement variance, data gaps and the block boundaries can affect this test. We quote $(n_{\rm exceed}+1)/(N+1)$ for the fixed-pair, rotation-path and catalog-contrast Monte Carlo tests. Conditional subset counts, source-scan frequencies and the finite set of catalog orientations are reported as empirical fractions. The 20 GV source-centered maximum-depth test and the date-influence tests recompute flux means for each shift. The six-bin positron count test holds the all-date bin means fixed and shifts all normalized positron bins together. In all cases the orbit stays fixed. The three shifted partitions are sensitivity checks on the null model.

For positions $a,b$ let $Z_k(a)=[D_k(a)-\mu_k(a)]/s_k(a)$, where $\mu_k$ and $s_k$ are the exact mean and standard deviation obtained by summing the independent block moments over all shifts. A pair uses $S_k=\min[Z_k(a),Z_k(b)]$. The target threshold for the original GC and Crab is $S_{20}=2.154$. Every scan compares the simulated maximum over its permitted positions and bins with this same \emph{observed target} threshold. With the nominal Bartels boundaries and 200\,000 trials the fixed-pair probabilities are 0.00435 in the target bin and 0.02385 across six bins. Using block phases 9 or 18 days changes the latter probabilities to 0.00560 and 0.02356. Independently simulated fixed-direction rows from the 30\,000-sky scan agree within Monte Carlo precision.

\begin{table}[!ht]
\caption{Conditional timing tails for a pair at least as strong as the target, using the same 30\,000 maps of date-shifted spectra. Each cell gives the count and the corresponding plus-one probability. The fixed-position row here uses fewer trials than the dedicated fixed-pair test.}
\centering\small\setlength{\tabcolsep}{4pt}
\begin{tabular}{lrr}
Allowed directions & 16.6-22.8 GV & Any of six bins\\\hline
True pair only & 136 (0.00457) & 731 (0.0244)\\
Either rotation path, photon-bright pairs & 254 (0.00850) & 1384 (0.0462)\\
Either rotation path, all eligible pairs & 11294 (0.3765) & 27604 (0.9201)\\
\end{tabular}

\end{table}

For the full catalog contrast the nominal time null uses 100\,000 trials, with tails 0.00269 at fixed rigidity and 0.01665 across six standardized bins. Using block phases 9 and 18 days yields six-bin tails of 0.0395 and 0.0685. Calendar-year shifts give 0.0342. None of these tests covers the full a posteriori selection.

\section{Rigid rotations of the two apertures}
Each rigid scan rotates the two centers, both footprints and their half-window boundaries around one pole in $0.5^\circ$ steps. There are 501 eligible orientations around the Galactic pole and 720 around the equatorial pole. The brightness restriction requires both ordered photon fluxes (fainter and brighter) to reach their observed counterparts. Interchanging the two windows gives the same set. Four distinct orientations pass. They comprise the true position shared by both scans, Galactic longitude shifts $+2.5^\circ$ and $+3.0^\circ$, and a right ascension shift $+0.5^\circ$. All overlap the original pair. At $+146^\circ$ in right ascension, the strongest alternative at the target bin according to $S_k$, the coordinates are $(225.37^\circ,-55.21^\circ)$ and $(32.42^\circ,56.29^\circ)$. The window point-source fluxes are $1.484\times10^{-10}$ and $4.200\times10^{-11}\,\mathrm{cm}^{-2}\,\mathrm{s}^{-1}$. PKS~0301$-$243 supplies 74.4\% of the flux in the rotated GC window. MS~1458.8+2249 supplies 48.4\% in the rotated Crab window. Their respective source-centered $V_+$ values have propagated correlations 0.654 and 0.773 with the $V_+$ of those rotated windows. A photon-fainter window can still contain a catalog source.

\begin{figure}[!ht]
\includegraphics[width=.94\textwidth]{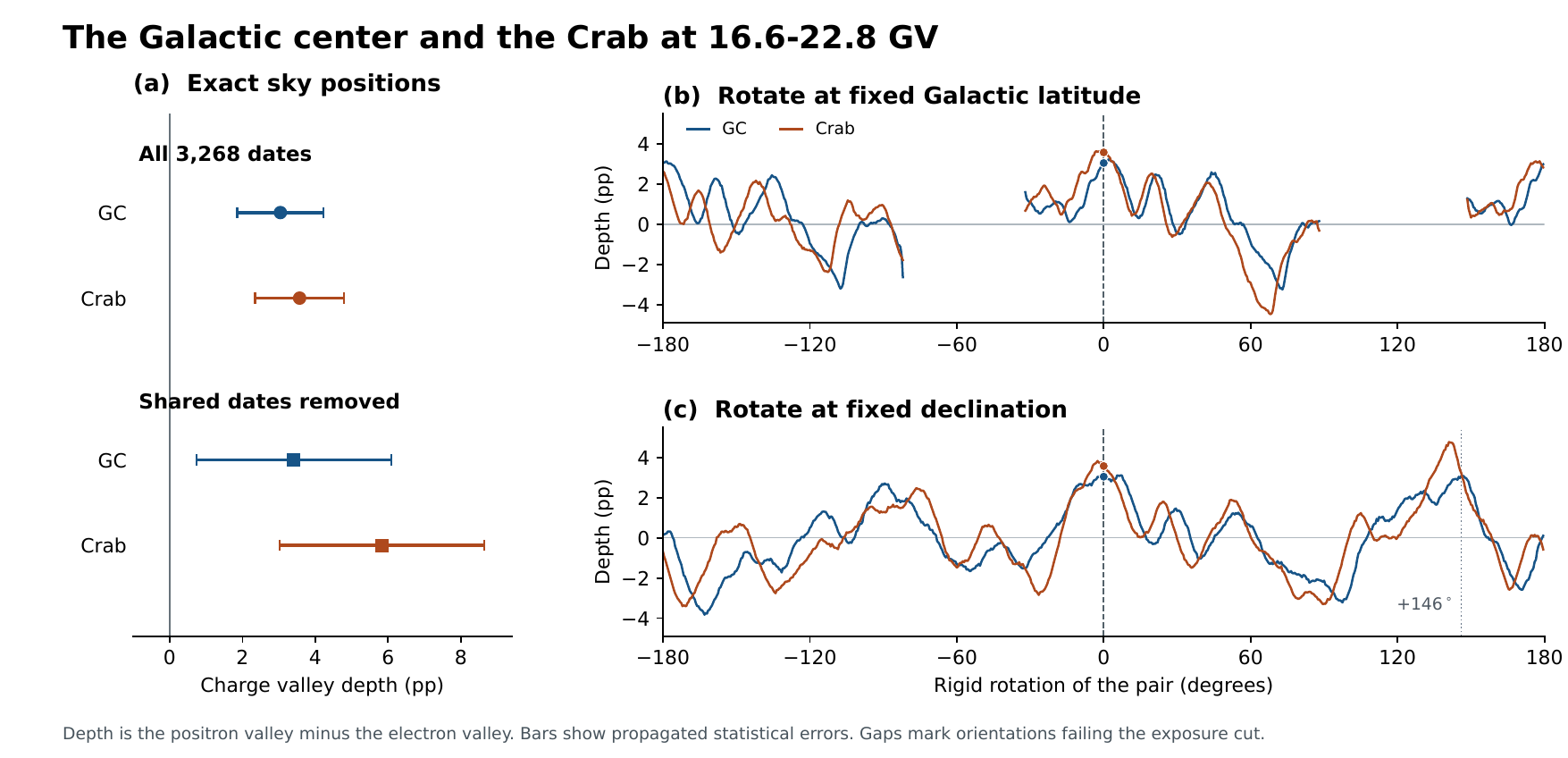}
\caption{Full-sample and shared-date-deletion estimates (left), and charge depths for each window while the pair is rotated along the two paths (right). Each rotation preserves the separation and transports both half-window boundaries. The dotted line marks the alternative pair at $+146^\circ$. Gaps indicate orientations that fail the original exposure requirement.}
\end{figure}

\clearpage
\section{Time distribution and influence of individual dates}
Here $q_{20,d}$ denotes the daily depth for the 16.6-22.8 GV bin. The contribution $c_{ad}=(w_{da}-g_d)q_{20,d}$ is signed and additive with the full-sample means and weights held fixed. The 350 common dates contribute 2.32 pp at the GC and 2.63 pp at the Crab. Of these totals, 2.12 and 2.26 pp accrue in 2012-2015 (142 shared days). Signed annual totals include cancellations. The all-date livetime-weighted rms of $q_{20,d}$ is 24.48 pp. Propagation of the independent tabulated statistical errors with fixed normalization means gives an rms of 24.23 pp. These marginal rms values do not determine the signed projection onto the orbital date weights. Each deletion below recomputes flux means, both half-window weights, and propagated statistical errors. Omitting any one of 136 populated Bartels blocks gives ranges 2.64-3.28 and 3.25-3.81 pp. Scanning all consecutive ten-calendar-day intervals gives minimum retained depths of 2.67 and 3.23 pp. Deleting the 20 largest $|c_{ad}|$ dates separately for each window leaves $1.29\pm1.22$ and $1.93\pm1.26$ pp. This last operation selects the omissions using the same data. It measures influence and has no discovery probability interpretation. The periods before and from January 1, 2016 contain 1598 and 1670 dates. Recomputing the reference within each period gives $D_{\rm GC}=5.38\pm1.69$ and $0.76\pm1.66$ pp, and $D_{\rm Crab}=5.17\pm1.75$ and $2.08\pm1.71$ pp. The period differences are 1.95 and 1.26 statistical standard errors, respectively.

We calibrated this influence measure at the two fixed directions and the target rigidity. For each block origin, one million skies use synchronized shifts of all six fluxes within their observed-date blocks. Every trial recalculates all six reference fluxes. We keep trials whose untrimmed $\min(Z_{\rm GC},Z_{\rm Crab})$ meets the data threshold, standardizing using the simulated means and dispersions for that origin. Separately for each window in each retained sky, we rank $|c_{ad}|$, omit the first 20 dates, and recompute both window weights and the reference. Our statistic is the mean of the two losses in $D$, with the same procedure applied to the data. The observed loss is 1.70 pp. With the nominal block origin, the conditional median loss is 0.71 pp and 96 of 4417 qualifying skies (2.17\%) lose at least 1.70 pp. The two other block phases give 113/1045 (10.81\%) for $m=9$ and 89/3860 (2.31\%) for $m=18$. These fractions describe the concentration in individual dates for an already deep projected pair. The post hoc deletions supply no independent significance for the pair or its position. A separate 100\,000-sky calculation yields 458 qualifying untrimmed pairs with reestimated flux normalizations and 457 with fixed normalizations on the same shifts.

\begin{center}
\includegraphics[width=.73\textwidth]{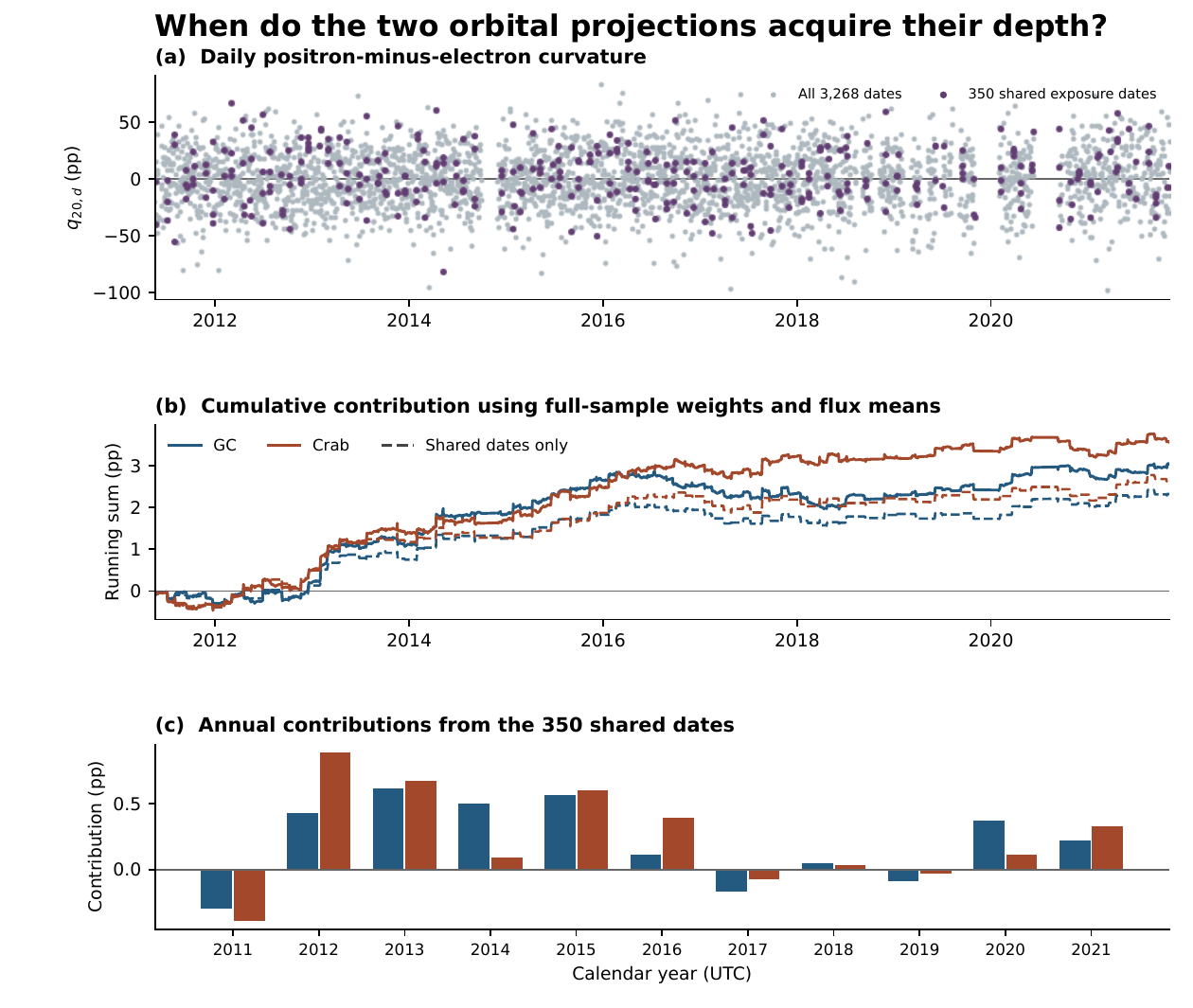}
\end{center}
\noindent\textbf{FIG. S3.} Daily charge curvature, its weighted cumulative contribution to each aperture, and annual contributions of the shared dates (pp). The first and last calendar years have partial coverage.

\clearpage
\section{Checks against external time variables}
We flag an observed UTC day if it overlaps the interval from the cataloged near-Earth interplanetary coronal mass ejection (ICME) disturbance to the ejecta end, including the sheath (or from ejecta start if the disturbance time is missing). We separately flag dates with any three-hour planetary $Kp\geq5$ or with an Oulu neutron-monitor daily count at least 3\% below the median of the preceding seven calendar days. The latter is an empirical ground-based proxy for a cosmic-ray decrease. The event list is the Richardson-Cane near-Earth ICME catalog, \url{https://izw1.caltech.edu/ACE/ASC/DATA/level3/icmetable2.htm}. Kp is from the GFZ index series (doi:10.5880/Kp.0001). We acknowledge the NMDB database (\url{https://www.nmdb.eu}), founded under EU FP7 contract 213007, and the Oulu neutron monitor, Sodankyl\"a Geophysical Observatory, University of Oulu, Finland, for its corrected count rates. All three indicators cover the matched AMS dates.

\begin{table}[h]
\caption{Stability checks at 16.6-22.8 GV. Entries are $D$ in pp with propagated statistical errors. The last row changes the reference weights.}
\centering\small\setlength{\tabcolsep}{4pt}
\begin{tabular}{lrrr}
Selection & Dates & GC & Crab\\\hline
All dates & 3268 & $3.04\pm1.19$ & $3.57\pm1.22$\\
Exclude ICME and sheath days & 2899 & $2.61\pm1.24$ & $3.43\pm1.29$\\
Exclude $Kp\geq5$ days & 3034 & $2.66\pm1.22$ & $3.40\pm1.25$\\
Exclude Oulu decreases & 3229 & $2.92\pm1.19$ & $3.47\pm1.23$\\
Exclude any of the three flags & 2748 & $2.35\pm1.27$ & $3.17\pm1.31$\\
Match reference in $10^\circ$ solar $\beta$ bins & 3268 & $2.90\pm1.16$ & $3.43\pm1.20$\\
\end{tabular}

\end{table}

For the largest 20 absolute contributions to the GC and Crab sums, respectively 3 and 5 dates overlap cataloged disturbances. One date on each list has $Kp\geq5$, and none meets the Oulu decrease criterion. The lists share 11 dates. Nineteen of the 20 GC dates and all 20 Crab dates belong to the 350 dates with exposure in both windows. The GC list spans 2011-2021 and the Crab list 2011-2017. These diagnostics do not identify the origin of a particular daily value.

We estimate the signed solar $\beta$ angle from the Sun's geocentric direction at UTC noon and the daily orbital angular-momentum unit vector, obtained from successive inertial position directions. For each target, match the total reference weight in each signed $\beta$ bin to the target weight, retaining relative reference livetimes within the bin and renormalizing each charge and rigidity separately. Bin widths of $5^\circ$, $10^\circ$ and $20^\circ$ give depth ranges 2.72-2.92 pp (GC) and 3.30-3.43 pp (Crab). This checks a dependence on daily orbital illumination. Thermal memory, rigidity-dependent acceptance, instrument settings and intra-day changes require further information.

\clearpage
\section{Individual central-bin deviations and in-flight annihilation}
\noindent Figure S4 shows the projected flux deviation at 16.6-22.8 GV before forming a difference across rigidity or charge. At the exact target coordinates the central positron deviations are $-3.49\%$ (GC) and $-3.45\%$ (Crab). The corresponding electron deviations are $+0.20\%$ and $+0.28\%$. Subtracting the central-bin deviation from the interpolated sideband deviation, as in Eq.~(2) of the main text, gives the 16.6--22.8 GV positron and electron depth maps in Fig.~\ref{fig:sky}(b) of the main text and Fig.~S6(b), respectively.
\begin{center}
\includegraphics[width=.97\textwidth]{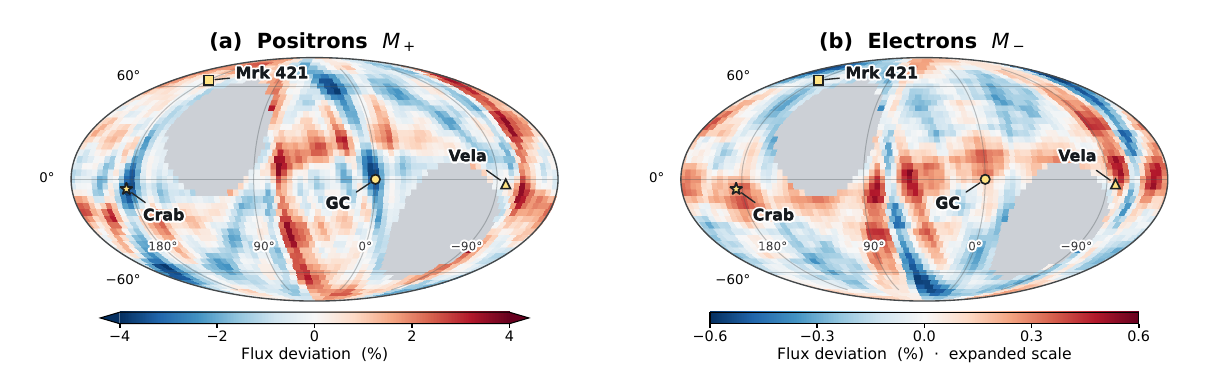}
\end{center}
\noindent\textbf{FIG. S4.} Projected central-bin $M_{sk}$ relative to the all-date mean for the same species and bin (percent). The positron color scale spans $\pm4\%$ and the expanded electron scale spans $\pm0.6\%$. The sign convention here gives negative values for below-reference flux.

\paragraph{Cold-electron column.} For a positron of unit charge at rigidity $R$, let $\gamma=\sqrt{1+[p/(m_ec)]^2}$ and $r_e$ be the classical electron radius. The exact two-photon cross section on an electron at rest [Krumholz, Crocker, and Sampson, Eq.~(66), cited in the main text] approaches $\pi r_e^2[\ln(2\gamma)-1]/\gamma$ at high energy. At 20 GV it is $6.545\times10^{-29}\,\mathrm{cm^2}$. For electron column $N_e$ and approximately constant rigidity, the annihilated fraction is $P(R)=1-\exp[-N_e\sigma(R)]$. As a benchmark, $10\,\mathrm{g\,cm^{-2}}$ of pure hydrogen gives $P(20\,\mathrm{GV})=3.91\times10^{-4}$. Absolute attenuation of 3\% (75\%) would require $N_e=4.65\times10^{26}$ ($2.12\times10^{28}$) $\mathrm{cm^{-2}}$. The 3\% is an illustrative attenuation of incident flux. The observed percentage-point depth also involves sidebands, orbital mixing and reference normalization.

To compare like observables, take the same column in both apertures and apply $P_k$ in all three bins. Use the geometric bin centers $13.51,19.45,30.91$ GV and a uniform incident sky, with the pencil-beam response $A,\bar f$ defined above under \emph{Finite angular deficits}. Then
\[
 M_{+k}(a)=-100\,\frac{(A_{a1}+A_{a2})P_k}{1-(\bar f_1+\bar f_2)P_k}.
\]
At this large column, the removed fractions are 0.861, 0.759 and 0.607 from low to high rigidity. Interpolating the projected sidebands gives $V_+=(0.0399,0.0425)$ pp for the GC and Crab. Applying a 75\% removal restricted to the central bin instead gives $(3.25,3.46)$ pp. This comparison isolates the energy dependence of direct annihilation at fixed column. It neglects energy losses in traversing matter and uses bin centers rather than folding a source spectrum with the instrument response. The very large column serves as a diagnostic of the proposed process.

For a bound over all common columns write $t=N_e\sigma(20\,\mathrm{GV})$, $F=\sum_j\bar f_j$, $a=\sum_j A_{aj}$ for the chosen window, and $y_i(t)=P_i(t)/(1-FP_i(t))$. Then $V_+=100a[y_M-(1-u)y_L-u y_H]$. Each $y_i$ increases with $t$. On any interval $[t_0,t_1]$, evaluating the positive term at $t_1$ and both subtracted terms at $t_0$ gives an upper bound. We tile $0\leq t\leq40$ in steps of $10^{-4}$. For the remaining tail use $y_M\leq1/(1-F)$ and both subtracted terms at 40. Evaluating the curve at $t=2.62533$ and enclosing its maximum by these interval bounds gives $0.08937$-$0.08941$ pp at the GC and $0.09508$-$0.09513$ pp at the Crab (bounds rounded outward). This result assumes the same cold-electron column in the two apertures, a pencil beam, fixed rigidity through the column, and the stated bin centers. The bound refers to the curvature of this forward model.
\clearpage
\section{Charge difference and three additional photon-bright regions}
\noindent Figure S5 shows the difference between the two lepton depths. From visual inspection of the lepton and photon maps we selected three regions near $(l,b)=(20^\circ,+46^\circ)$, $(138^\circ,-22^\circ)$ and $(22^\circ,-56^\circ)$. In each we chose the eligible aperture with the largest 3FHL photon flux within $6^\circ$ of the stated coordinates. Radii of $8^\circ$ and $10^\circ$ select the same centers. All depths use the full sample.
\begin{center}
\includegraphics[width=.60\textwidth]{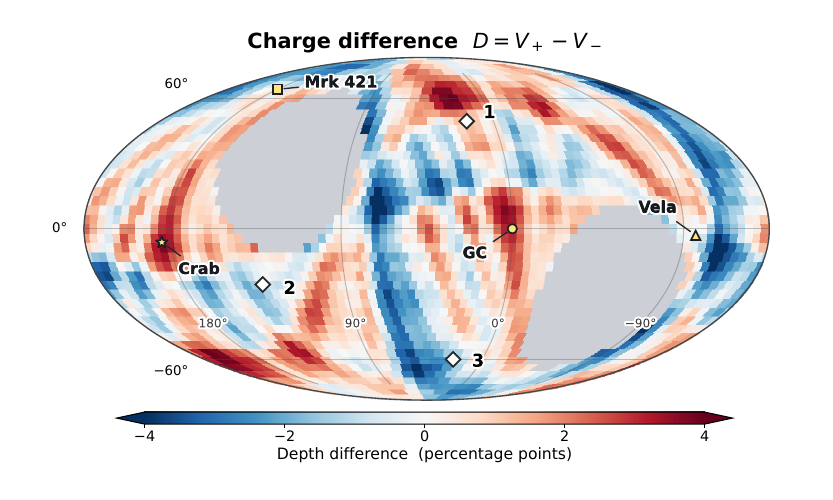}
\end{center}
\noindent\textbf{FIG. S5.} Projected charge difference $D=V_+-V_-$ at 16.6-22.8 GV. The scale spans $\pm4$ pp. White diamonds mark the three aperture centers listed below. Yellow symbols mark the four reference positions in Table I of the main text.

\begin{table}[h]
\caption{Three visually selected photon-bright apertures. $G$ is the summed 16.6-22.8 GeV point-source photon flux in units of $10^{-10}\,\mathrm{cm}^{-2}\,\mathrm{s}^{-1}$. The spectral depths are in percentage points with propagated statistical errors.}
\centering\small\setlength{\tabcolsep}{4pt}
\begin{tabular}{crrrrr}
Region & $(l,b)$ [deg] & $G$ & $V_+$ & $V_-$ & $D$\\\hline
1 & $(18.0,+48)$   & 5.91 & $+1.49\pm1.28$ & $0.00\pm0.17$ & $+1.49\pm1.30$\\
2 & $(136.1,-24)$  & 4.83 & $-1.08\pm0.94$ & $-0.13\pm0.12$ & $-0.95\pm0.95$\\
3 & $(23.5,-60)$   & 4.94 & $-2.18\pm1.10$ & $-0.06\pm0.14$ & $-2.12\pm1.11$\\
\end{tabular}

\end{table}

The positron central bin lies below its interpolated sidebands in region 1 and above them in regions 2 and 3. The regions were selected retrospectively. We matched the catalog positions of all pointlike 3FHL entries to Eq.~(1), finding 5, 6 and 8 entries in regions 1-3. Table S4 gives every entry that supplies at least 30\% of its region's fitted photon flux. Regions 1 and 3 each have one dominant source. Two sources supply 94.7\% in region 2.

\begin{table}[h]
\caption{Contributors to the bright apertures in Table S3. $f_\gamma$ is the fraction of the aperture's fitted photon flux at 16.6-22.8 GeV. $\theta$ is the angular distance from its grid center to the 3FHL position. The last column gives the positron depth after centering the same aperture on the source, with its statistical error.}
\centering\small\setlength{\tabcolsep}{4pt}
\begin{tabular}{cllrrr}
Region & 3FHL source & Counterpart & $f_\gamma$ [\%] & $\theta$ [deg] & Source-centered $V_+$ [pp]\\\hline
1 & J1555.7+1111 & PG 1553+113    & 85.3 & 4.87 & $+0.90\pm1.26$\\
2 & J0222.6+4302 & 3C 66A         & 59.7 & 8.16 & $-0.80\pm0.74$\\
2 & J0136.5+3906 & B3 0133+388   & 35.0 & 3.54 & $-0.83\pm0.88$\\
3 & J2158.8$-$3013 & PKS 2155$-$304 & 86.1 & 8.38 & $-0.06\pm1.09$\\
\end{tabular}

\end{table}
All four entries are detected LAT sources. In 3FHL the BL Lac counterparts of PG~1553+113 and 3C~66A are classified as identified (\texttt{BLL}), and those of B3~0133+388 and PKS~2155$-$304 as associated (\texttt{bll}). PKS~2155$-$304 has also been directly observed in a joint LAT and H.E.S.S. campaign [F.~Aharonian \textit{et al.}, Astrophys. J. Lett. \textbf{696}, L150 (2009)]. Each source-centered aperture passes the exposure cut in both halves. For PKS~2155$-$304 the change from the grid center to the source position is $+2.12\pm1.08$ pp, including the shared daily-flux covariance. The measured depths in Table S4 do not establish a positron dip at any of these source positions.
\clearpage
\section{Scan at the positions of 3FHL point sources}
For each of the 1508 pointlike 3FHL entries, we center the rectangular aperture of Eq.~(1) on its catalog coordinates and compute the two half-window livetimes from the minute-sampled orbit. Each of 1177 entries meets the original requirement of 100 effective dates per half. We use all 3268 common dates for the reference. All entries are ranked by the measured $V_+$ at 16.6-22.8 GV, including positions with no local minimum. The first 11 entries have deeper point estimates than the Crab, which ranks 12th. Thirty-four are deeper than the mathematical GC.

\begin{table}[h]
\caption{Source-centered examples. The first two rows are the largest $V_+$ estimates in the complete scan. The last two are all catalog entries with identified counterparts (uppercase 3FHL class) outside the $18.20^\circ$ GC/Crab and antipode mask with $V_+$ above the GC value. $r_V$ ranks depth among the 1177 eligible entries. $r_\gamma$ ranks the individual source's fitted 16.6-22.8 GeV photon flux among all 1508 pointlike entries. $f_\gamma$ is the source's fraction of the photon flux in its own aperture. Depths and statistical errors are in pp.}
\centering\small\setlength{\tabcolsep}{4pt}
\begin{tabular}{llrrrrr}
3FHL source & Counterpart & $r_V$ & $r_\gamma$ & $f_\gamma$ [\%] & $V_+$ & $V_-$\\\hline
J1500.9+2238 & MS 1458.8+2249 & 1 & 269 & 5.4 & $4.12\pm1.17$ & $0.26\pm0.15$\\
J0252.8$-$2218 & PKS 0250$-$225 & 2 & 321 & 12.6 & $4.11\pm1.18$ & $0.23\pm0.16$\\
J0303.4$-$2407 & PKS 0301$-$243 & 6 & 30 & 73.5 & $3.97\pm1.15$ & $0.22\pm0.15$\\
J1427.0+2348 & PKS 1424+240 & 19 & 9 & 80.2 & $3.41\pm1.18$ & $0.17\pm0.15$\\
\end{tabular}

\end{table}

The first two centers lie $1.91^\circ$ from exact antipodes. Their positron depths have propagated statistical correlation $0.950$ from the shared daily flux measurements. The first center lies $7.89^\circ$ from PKS~1424+240, which supplies 81.7\% of the photons from catalog point sources in that aperture. The second lies $3.03^\circ$ from PKS~0301$-$243, which supplies 77.6\%. Each fraction uses the sum of fitted photon fluxes. Figure~\ref{fig:shape}(i) shows the spectrum at the deepest catalog center. Recentering on the two dominant photon sources gives the last two rows of the table and the spectra in Fig.~\ref{fig:shape}(j,k). Both are identified BL Lac objects with independent very-high-energy detections [A.~Abramowski \textit{et al.}, Astron. Astrophys. \textbf{559}, A136 (2013); V.~A.~Acciari \textit{et al.}, Astrophys. J. Lett. \textbf{708}, L100 (2010)].

For the fixed-rigidity time-null test, in each realization we cyclically shift the observed daily records within each 27-day block, with a common shift for all bins and source apertures. We recalculate the three flux reference means, project onto all 1177 positions and compare the largest $V_+$ with the observed 4.12054 pp. The nominal origin gives 1379 exceedances in 10\,000 trials (13.79\%, 95\% binomial interval 13.12-14.48\%). Using the common convention $\lfloor(n+m)/27\rfloor$, phases $m=9$ and 18 days give 1490/10\,000 (14.90\%) and 1421/10\,000 (14.21\%), respectively. This tail accounts for maximizing depth over source positions at the stated rigidity. The separate catalog-brightness and rotation tests in the main text address angular specificity.

\clearpage
\section{A directional test at PKS 0301$-$243 and PKS 1424+240}
The two identified photon sources in Fig.~\ref{fig:shape}(j,k) are $171.68^\circ$ apart. Their projections contain 503 and 494 dates, with 338 in common. The common dates carry 75.5\% and 78.5\% of their respective full-sample weights. Their full-sample $V_+$ estimates have statistical correlation 0.486 and their difference (0301 minus 1424) is $0.56\pm1.18$ pp. As a timing check we exclude the 338 dates from both numerators and the all-date reference, and recompute the normalized half-aperture weights and flux means. On the remaining 165 and 156 contributing dates we obtain $V_+=3.71\pm3.44$ and $5.12\pm4.05$ pp, respectively. These reduced samples give broad statistical intervals and reuse the dates used to select the sources. Giving the two windows identical daily weights would make their spectra identical by construction, because the published inputs supply a single flux per date, rigidity and charge.

A directional measurement can retain the same 2011-2021 dates, the original aperture geometry, and the three rigidity bins of the main text. For each source separately, one choice of control is the union of two apertures at the same declination displaced by $\pm30^\circ$ in right ascension, fixed before inspecting tracks. Bin primary leptons by asymptotic arrival direction after geomagnetic backtracking. Let $N_{\rm on}$ and $N_{\rm off}$ be the counts for one charge and rigidity in a common time interval, and let $E_{\rm on}$ and $E_{\rm off}$ be the corresponding exposures, including charge-dependent transmission, selection and acceptance. For a source-to-control flux ratio $r$ constant over the fitted intervals and an arbitrary common intensity $I_t$ in each interval, the expected counts are $r I_t E_{\rm on}$ and $I_t E_{\rm off}$. Under independent Poisson sampling of signal events,
\[
 N_{\rm on}\mid (N_{\rm on}+N_{\rm off}=n)
 \ \sim\ {\rm Binomial}\!\left(n,\frac{r E_{\rm on}}{r E_{\rm on}+E_{\rm off}}\right).
\]
The unknown $I_t$ cancels separately for each charge and rigidity. An interval with exposure on only one side carries no information about $r$. Residual proton background, charge confusion, rigidity migration and their uncertainties must enter the count model for real data. Counts and exposures binned this way, together with the relevant response and background information, would suffice for a directional test. The public AMS tables supply one flux per date, charge and rigidity. NASA AMS02RATES supplies one-second all-particle rates and orbital information. The AMS data management plan describes the collaboration distribution of reconstructed events [\url{https://ams.cern.ch/AMS-Data-Management-Principles-2017.pdf}].

Fit $r_{s,L},r_{s,M},r_{s,H}$ for each charge, and form $C_s=100[(1-u)r_{s,L}+u r_{s,H}-r_{s,M}]$ with the same $u$ as in the main text. Report $C_+$ and $C_+-C_-$ with their joint errors at each target. The proposed two-source pattern gives $C_+>0$ and $C_+-C_->0$ at each position. Also report $r_{+,L}-r_{+,M}$ and $r_{+,H}-r_{+,M}$ with their covariance. Positive steps on both sides distinguish a local minimum from curvature in a monotone spectrum. The daily projection does not predict a track-level amplitude. A test on the original dates checks localization using the same underlying events. Independent confirmation would require separate data.

\clearpage
\section{Six-bin positron count and electron controls}
We evaluated the same positron-only selection at all 1177 eligible 3FHL centers. For each of the six interior rigidity bins, the central $M_{+k}$ must lie below both neighbors and $V_{+k}$ must reach the full-precision GC depth measured in the 16.6--22.8 GV bin (3.029 pp rounded). This cut was fixed across bins after inspecting the GC. The centers are overlapping apertures and share dates. Each passing center contributes one count to its rigidity bin.

\begin{center}
\includegraphics[width=.97\textwidth]{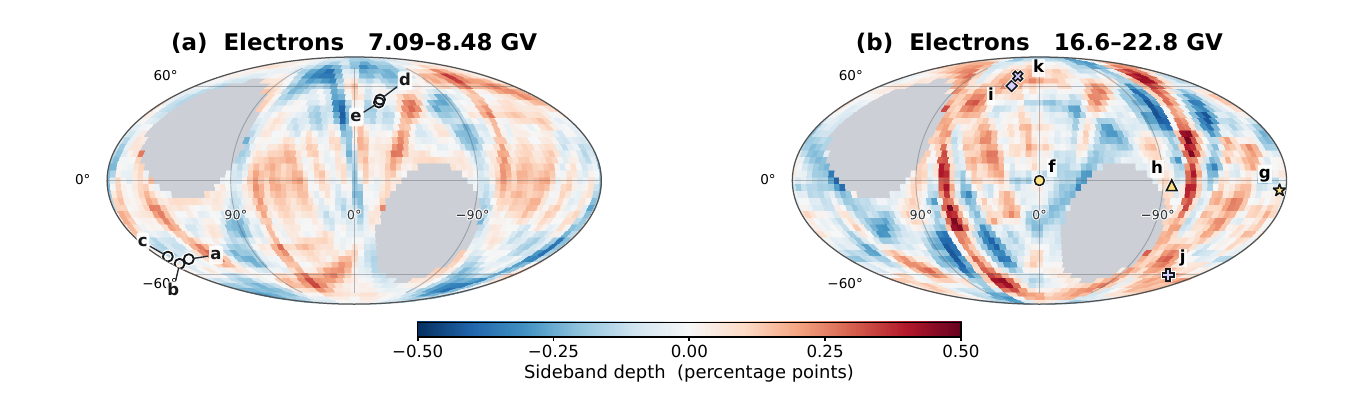}
\end{center}
\noindent\textbf{FIG. S6.} Electron sideband depths $V_-$ at (a) 7.09--8.48 and (b) 16.6--22.8 GV. Both panels span $\pm0.5$ pp. The corresponding positron maps in Fig.~\ref{fig:sky} span $\pm5$ pp. Each map uses the same dates, aperture weights, sidebands, and exposure mask as its positron counterpart. Letters identify the positions in Fig.~\ref{fig:shape} of the main text.

\medskip
\noindent The bins 4.02--5.90, 5.90--7.09, 7.09--8.48, 8.48--11.0, 11.0--16.6 and 16.6--22.8 GV contain respectively 0, 0, 5, 0, 0 and 34 selected centers. The lower-band centers correspond to ZS~0214+083, NGC~1068, PKS~B0253+033, NVSS~J141133$-$072252 and NVSS~J141922$-$083830. The median positron depths are 3.28 pp in the lower group and 3.46 pp in the upper group. No center belongs to both groups. Figure~\ref{fig:shape}(a--e) of the main text shows the positron and electron profiles for all five lower-band windows.

In each of 4,000 trials we draw a circular date shift independently for each 27-day Bartels block, apply it to the spectra at all rigidities, and recompute the six counts with the original orbital weights. At the 3.029 pp cut, 104 shifts give at least five lower-bin windows and 83 give at least 34 upper-bin windows. Including the observation and ties gives fixed-bin upper tails of $105/4001=0.0262$ and $84/4001=0.0210$. Allowing any of the six bins to attain so small a binwise tail gives $475/4001=0.1187$ and $363/4001=0.0907$, respectively.

None of the 4,000 shifts meets both count thresholds in the same trial. For the bin pair, counts and cut identified from the data, let $q$ denote the probability that another shift meets both thresholds. Zero successes in 4,000 independent draws gives the one-sided 95\% Monte Carlo upper bound $1-0.05^{1/4000}\simeq 0.00075$ on $q$. The following test allows the bin pair and cut to vary.

\paragraph{Recurrence in any two bins.} For a depth cut $c$, let $N_{ik}(c)$ be the number of passing windows in rigidity bin $k$ for the observation ($i=0$) or one of the 4,000 shifted skies. We compute the upper-tail rank in each bin by including all ties,
\[
 u_{ik}(c)=\frac{\#\{j:N_{jk}(c)\ge N_{ik}(c)\}}{4001},
 \qquad u_{ik}(c)=1\quad\text{if }N_{ik}(c)<2 .
\]
The second-smallest of the six $u_{ik}(c)$ defines $R_i(c)$. A low value requires large counts relative to each bin's own null distribution in at least two bins. The other four counts may take any value. At the GC-derived cut the two smallest observed tails are $84/4001$ and $105/4001$. For 24 shifted skies the second tail is at least as small as the observed $105/4001$, giving $p=25/4001=0.00625$ with the observation included (one-sided Gaussian equivalent $2.50\sigma$).

To allow for the choice of depth cut, we repeat the calculation at $c=2.00,2.25,\ldots,4.00$ pp. The GC-derived cut is evaluated separately and adds no grid point to this scan. At each cut we rank $R_i(c)$ against all 4,001 values to obtain $p_i(c)=\#\{j:R_j(c)\le R_i(c)\}/4001$. We then take the smallest $p_i(c)$ over the nine cuts for each sky and rank these minima. The observed minimum occurs at 3.25 pp, where three lower-band and 23 upper-band windows pass. Of the 4,000 shifts, 80 match or exceed that result, giving $p=81/4001=0.02024$ (one-sided $2.05\sigma$, Monte Carlo standard error $0.0022$). Moving the first block boundary later by $\Delta=9$ and 18 days, with block index $\lfloor(n-\Delta)/27\rfloor$, gives $93/4001$ and $80/4001$, respectively. The calibration includes the choices of bin pair and cut within this scan. The catalog, aperture geometry, local-minimum definition and choice to test recurrence are held fixed.

\clearpage
\section{Fermi-LAT photon counts and the 3FHL flux template}
\begin{center}
\includegraphics[width=.76\textwidth]{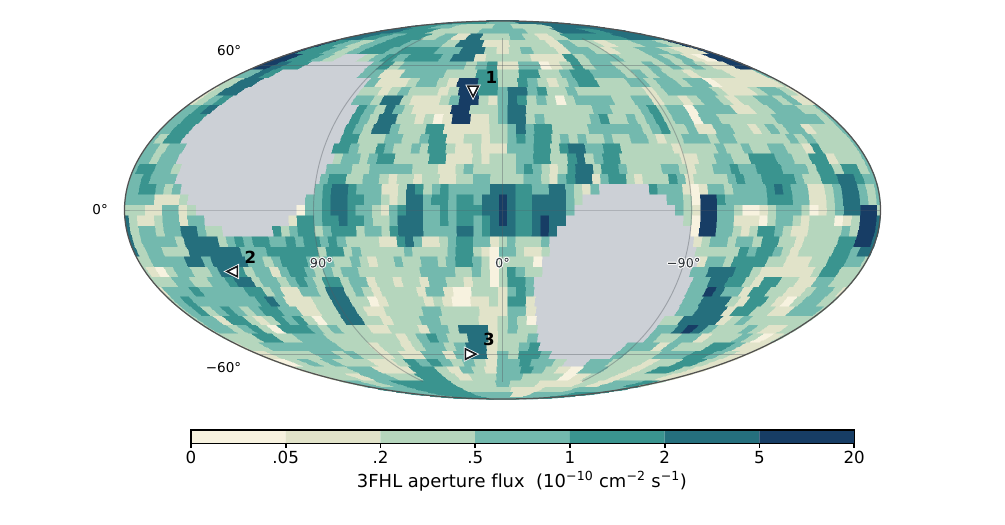}
\end{center}
\noindent\textbf{FIG. S7.} Modeled 3FHL point-source flux summed within each aperture at 16.6--22.8 GeV. The numbered triangles mark the photon-bright apertures in Table~I of the main text. Gray marks cells outside the AMS exposure cut used for this comparison. Source brightness in the main-text contrast is taken from this template. The broadband photon counts in Fig.~\ref{fig:sky}(c) of the main text cover the full sky.

\medskip
\noindent For the photon-count map in Fig.~\ref{fig:sky}(c) we used the public LAT weekly Pass~8 SOURCE event files (class mask 128) and spacecraft histories. We retained both conversion types, $E_\gamma\geq100$ MeV, zenith angle $<90^\circ$, photon good-time intervals, and spacecraft intervals with \texttt{DATA\_QUAL}$>0$ and \texttt{LAT\_CONFIG}$=1$. Event times were converted to the AMS UTC dates. In total 155,919,993 events fall on 3247 dates. The other 21 AMS dates fall during the March--April 2018 LAT observing interruption. Each event contributes to one Galactic HEALPix pixel ($N_{\rm side}=128$, RING ordering). The map includes the variation in LAT exposure. The point-source photon fluxes, aperture rankings and rotation tests use the 3FHL template in Fig.~S7.

\paragraph{Positional overlap on the AMS grid.} We applied the six-bin positron selection at the 3.029 pp cut to the 2,094 exposure-eligible grid centers before matching to the catalog. The counts in ascending rigidity are 0, 0, 8, 0, 0 and 50. For each selected aperture we counted a match if it contained at least one of the 1,508 pointlike 3FHL entries, using the original local window boundaries of Eq.~(1). With the AMS map fixed, we rotated the entire catalog about the Galactic pole through $0^\circ,1^\circ,\ldots,359^\circ$. This preserves source latitudes and separations and retains the overlap among apertures. The angular rank includes the observed orientation and ties.

At zero rotation all 8 lower-bin and all 50 upper-bin apertures match. A complete match occurs in 342 of the 360 orientations for the lower bin, 226 for the upper bin, and 216 for both together. As a complementary measure of angular proximity, the median distance from a selected grid center to its nearest catalog source is $2.25^\circ$ across the 58 apertures. The median over orientations is $2.47^\circ$, and 81 of 360 orientations give a median distance as small as observed. These comparisons condition on the measured dip map, depth cut and catalog latitude distribution.

\paragraph{Flux-weighted rotation of the catalog.} At the same fixed dip apertures, the mean 16.6--22.8 GeV catalog flux per aperture is $0.973$ and $3.066$ in the lower and upper rigidity groups ($10^{-10}\,{\rm cm}^{-2}\,{\rm s}^{-1}$). The medians over rotations are $0.653$ and $0.816$. Of the 360 orientations, 85 and 7, respectively, give group means at least as high as observed. Pooling all 58 apertures counts a source again whenever it appears in another aperture. The mean is then $2.777$ (rotation median $0.834$), reached or exceeded in 6 of 360 orientations. For the angular comparison in the main text, we sum each source's flux once over the union of all 58 apertures. The observed total is $4.30\times10^{-9}\,{\rm cm}^{-2}\,{\rm s}^{-1}$. Seven of 360 orientations reach or exceed it. They are the actual catalog orientation and six offsets between $177^\circ$ and $183^\circ$. This rank conditions on the observed dip apertures, the GC-derived cut and Galactic-longitude rotations.

\clearpage
\section{Flux-weighted dip maps under date shifts}
We hold the 3FHL point-source positions and fluxes $F_s$ fixed and reselect the apertures on the observed and date-shifted maps. Each realization uses the same block shifts at all 2,094 eligible AMS grid centers and all six interior rigidity bins (Sec.~12). At cut $c$, let $A_{ik}(c)$ be the set of apertures with a qualifying minimum in realization $i$ and bin $k$. We form the set of catalog sources $U_{ik}(c)$ inside their union and evaluate
\[
 W_{ik}(c)=\sum_{s\in U_{ik}(c)} F_s,\qquad
 P_i(c)=\sum_{s\in\bigcup_k U_{ik}(c)} F_s .
\]
Each distinct catalog entry contributes once per bin to $W$ and once across the entire sky and all six bins to $P$. All entries use the same 16.6--22.8 GeV photon-flux template of Fig.~S7. The dip depth determines aperture selection. At 3.029 pp, $W$ is $6.21\times10^{-10}$ and $3.68\times10^{-9}\,{\rm cm}^{-2}\,{\rm s}^{-1}$ in the two populated bins. Their 31 and 102 entries are disjoint, giving $P=4.30\times10^{-9}\,{\rm cm}^{-2}\,{\rm s}^{-1}$. The Crab lies in five passing upper-bin apertures and contributes once to $P$.

For the pooled test we rank $P_i(c)$ by its upper tail among all 4,001 maps. For two-bin recurrence we rank $W_{ik}(c)$ within each bin, assign unit tail to bins containing fewer than two qualifying apertures, and rank the second-smallest binwise tail in each map as in Sec.~12. For each test the threshold scan ranks each map's most extreme result over the same nine cuts. At 3.00 pp the observed distinct-source recurrence attains its smallest tail on that grid. Ties are included throughout. Scoring each entire realization together retains common dates, neighboring apertures, antipodal correlations and correlations between rigidity bins. The statistics depend on the extent of the selected sky as well as its photon brightness.

\begin{table}[h]
\centering
\caption{Date-shift realizations at least as extreme as the observed map, out of 4,000. The fixed cut is the 3.029 pp GC depth. The scan covers nine cuts from 2.00 to 4.00 pp. The Monte Carlo tail for an entry $n$ is $(n+1)/4001$. All rows use the same 2,094 AMS grid centers and the same shifts.}
\small
\begin{tabular}{lrrrr}
\hline
 & \multicolumn{2}{c}{All six bins pooled} & \multicolumn{2}{c}{Recurrence in any two bins}\\
Weight & Fixed cut & Nine cuts & Fixed cut & Nine cuts\\\hline
Number of passing apertures & 170 & 428 & 45 & 179\\
Flux summed over passing apertures & 7 & 11 & 16 & 42\\
Flux of distinct sources & 19 & 34 & 3 & 33\\
Flux of distinct sources, Crab omitted & 46 & 110 & 16 & 38\\\hline
\end{tabular}
\end{table}

The aperture-flux row counts a source in every passing aperture that contains it. The last row sets only the Crab catalog entry's flux to zero in both the observed and shifted skies. The apertures and all other sources remain eligible. Its nine-cut two-bin tail is $39/4001=0.0097$ (one-sided $2.34\sigma$). These exploratory tails calibrate the bin and threshold searches for the stated weighting and within-block time-shift hypothesis. The photon template and the choice of weighting were informed by the observed maps.

\paragraph{Reproducing the results.} The Short Communication reproducibility archive contains the observational analysis in \texttt{observations/} (v2.3) and the two-bin recurrence analysis in \texttt{recurrence/}. The original analysis reconstructs the matched dates, orbital weights, catalog fluxes, shared covariances, and 20 GV controls. The new analysis reproduces the catalog counts and their fixed-bin and six-bin timing tails, the spectra in Fig.~\ref{fig:shape}, the two positron maps in Fig.~\ref{fig:sky}, the photon counts in Fig.~\ref{fig:sky}(c) and the two electron maps in Fig.~S6. The recurrence directory in the analysis archive supplies the saved trial counts, rank check, and reduced inputs and code for regenerating the date shifts. The spatial-overlap archive records source membership in the 58 apertures. The flux-weighted analysis archive provides the grid inputs, distinct-source list, block offsets, trial scores, rotation results and rank checks. The raw photon reconstruction reads the listed weekly photon and spacecraft files. The ancillary numerical data include every flux deviation and statistical error in Fig.~\ref{fig:shape}, the gridded depths for both charges and both rigidity bins, per-center six-bin depths and selections, the HEALPix counts, and the daily photon counts.

\end{document}